\documentclass{osa-article}

\journal{osac}

\articletype{Research Article}

\begin{document}

\title{Efficient broadband THz generation in BNA organic crystal at
Ytterbium laser wavelength}

\author{Hovan Lee,\authormark{1,*} Claudia Gollner,\authormark{2} Jiaqi Nie,\authormark{4}, Yan Zhang,\authormark{4} Cedric Weber,\authormark{1} and Mostafa Shalaby\authormark{3,4,$\dagger$}}

\address{\authormark{1}Department of Physics, Faculty of Natural \& Mathematical Sciences, King's College London, London, WC2R2LS, UK\\
\authormark{2}Photonics Institute, TU Wien, Gusshausstrasse 27-387, A-1040 Vienna, Austria\\
\authormark{3}Swiss Terahertz Research-Zurich, Techpark, 8005 Zurich, Switzerland and Park Innovaare, 5234 Villigen, Switzerland\\
\authormark{4}Key Laboratory of Terahertz Optoelectronics, Beijing Advanced Innovation Center for Imaging Technology Capital Normal University, Beijing 100048, China}

\email{\authormark{*}hovan.lee@kcl.ac.uk} 
\email{\authormark{$\dagger$}most.shalaby@gmail.com}


\begin{abstract}
In this work, we demonstrate BNA’s high potential for efficient generation of high power THz using ytterbium laser wavelengths. We study the generation theoretically and experimentally using laser wavelength of $960-1150$ $nm$. Broadband pulses of $0 - 7$ $THz$ and high efficiency of $0.6\%$ are demonstrated.
\end{abstract}

\section{Introduction}
Throughout the past decades, THz technology ($0.1-10$ $THz$) have proven invaluable in numerous disciplines \cite{THzSpectroscopyMasayoshi:2007, THzSpectroscopyJepsen:2011,THzSpectroscopyBaxter:2011} ranging from ultrafast magnetisation \cite{ApplicationBaierl:2016,mostafa_ni,thz,thz2} to biomedical sciences  \cite{ApplicationYang:2016,biomed1,biomed2,biomed3}, amongst others \cite{ApplicationUlbricht:2011,ApplicationRazzari:2009,ApplicationGreenland:2010,ApplicationShalaby:2017,ApplicationKolner:2008,ApplicationMcIntosh}. However, one major hurdle exists across all THz applications: The lack of convenience in generating THz radiation.

Regarding this problem, optical rectification (OR) of nonlinear crystals offers a suitable solution, as this method offers the portability and ease-of-use as a tabletop THz source, whilst maintaining a large optical to THz conversion efficiency. One advantage of this method comes from the freedom of choice in crystals, where the applicable pump wavelength range and THz spectrum can be changed according to the requirements of the situation.

In OR, the nonlinear crystal is pumped with a femtosecond (fs) electromagnetic pulse,
and a frequency dependant polarisation is induced within the crystal. Due the wide spectral contents of the fs pulse, the different frequency components superpose in the form of a beating polarisation, leading to emissions in the THz range. 

Different
crystals respond to pump pulses differently, generating characteristic THz spectra. In this regard, Lithium Niobate is a popular crystal for THz generation \cite{LiNbZhang:2020,LN3,LN4,LN1}. However, the spectrum is mainly limited to sub 1 THz and the generation geometry is complex (non collinear) to achiever phase matching.

In this sense, DAST \cite{DASTHauri:2011,DASTVicario:2015} and its derivative DSTMS \cite{DSTMSVicario:2014}, have been shown to generate wide THz spectra up towards 20 THz, with favourable phase matching conditions when paired with a pump pulse of wavelength $\sim 1.5$ $\mu m$.
This presents another problem, as the availability of conventional lasers at this spectral range is limited; most groups rely upon Optical Parametric Amplifiers (OPAs) in order to convert Ti:Sa ($800$ $nm$) or Yb ($1030$ $nm$) lasers to longer wavelengths.

Conversely, BNA (N-benzyl-2-methyl-4-nitroaniline) features phase matching at optical wavelengths below $1200$ $nm$, alleviating the shortcomings of DAST. In support of this claim, past reports on BNA demonstrated $0.2 - 3$ $THz$ spectra at $0.2\%$ conversion efficiency and $0.2 - 7$ $THz$ spectra at $0.8\%$ efficiency when pumped with $800$ $nm$ Ti:Sa \cite{BNAShalaby:2016} and $1200- 1120$ $nm$ near infrared \cite{BNA_NIRShalaby:2019} respectively.

In this work, we demonstrate BNA's high potential of THz generation by providing experimental data and theoretical analysis of the spectra generated under the more and more conventional ytterbium laser wavelengths of $960-1150$ $nm$ with varied crystal thicknesses. 

\section{Method}
\subsection{Experimental Setup}
The ytterbium laser was not available to us. Therefore, to achieve the specified pump wavelengths, we used a setup with similar wavelengths and higher tunability based on a 1kHz OPA equipped with a difference frequency generation (DFG) unit, pumped by a $3.5\ \mathrm{mJ}$, $800\ \mathrm{nm}$, $35\ \mathrm{fs}$ laser. The OPA beam is then sent to the BNA crystal (with variable thicknesses of $300\ \mathrm{\mu m}, 550\ \mathrm{\mu m}, 1100\ \mathrm{\mu m}$, from SwissTHz) through which the THz spectra are collinearly generated. The THz spectra are then refined through three low pass filters (LPF) with out-of-band rejection ratio of $>0.1\%$ to filter out the residual NIR frequencies. Subsequently the filtered THz spectra are expanded and focused with a set of three off axis mirrors, and detected with either a Golay cell (MTI Inc.), an electro-optical sampler ($100\ \mathrm{\mu m}$ GaP, from SwizzTHz), or a RIGI uncooled mircobolometer camera (from SwissTHz).

\begin{figure}
    \centering
    \includegraphics[width=0.6\linewidth]{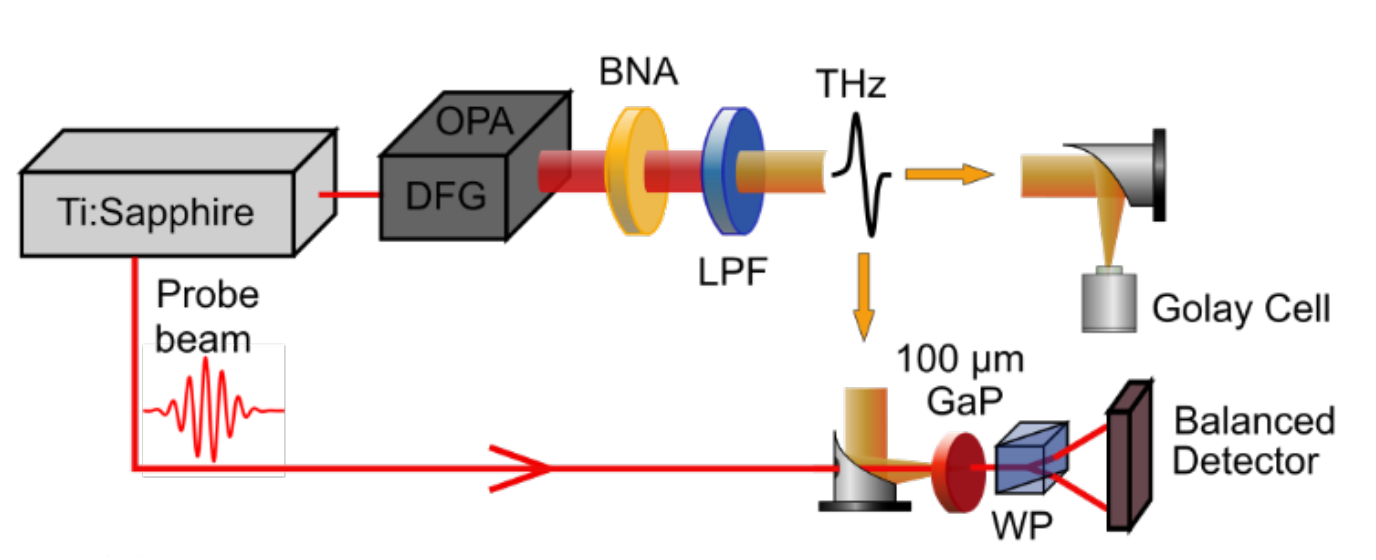}
    \caption{Schematic diagram of the experimental setup.}
    \label{setup}
\end{figure}

\subsection{Theoretical Spectra Calculation}
The simulated THz spectra and phase matching conditions were calculated by following the OR framework of Schneider Et Al. \cite{theory}; through considering the solution to the nonlinear wave equation inside BNA crystals:

\begin{equation}
    -\nabla\times\nabla\times\mathbf{E}(\mathbf{r},\omega)+\omega^2\mu_0\epsilon_0\epsilon(\omega)\otimes\mathbf{E}(\mathbf{r},\omega)-i\omega\mu_0\sigma(\omega)\otimes\mathbf{E}(\mathbf{r},\omega)=-\omega^2\mu_0\mathbf{P}_{NL}(\mathbf{r},\omega)
\end{equation}

Where $\mathbf{E}$ is the electric field, $\epsilon$ is the dielectric tensor, $\sigma$ is the optical conductivity tensor, and $\mathbf{P}_{NL}$ is the nonlinear polarisation (and can be expanded in powers of $\mathbf{E}$), all evaluated at frequency $\omega$ and position $\mathbf{r}$. Taking only the quadratic term in $\mathbf{P}_{NL}$:

\begin{equation}
    P_{NL,i}(\omega)=\epsilon_0\int d\omega' \chi^{(2)}_{ijk}(\omega;\omega',\omega-\omega')E_j(\omega')E_k(\omega-\omega')
\end{equation}

With $P_{NL,i}$ as the nonlinear polarisation in the $i$ direction, and $\chi^{(2)}_{ijk}$ is the 2nd order susceptibility of the material. The Electric field of the laser pulse may be characterised as a single axis polarised, time varying, complex amplitude $E_{0,k}(t)$ propagating with carrier frequency $\omega_0$.
This simplifies the nonlinear response by decreasing the number of active elements of the susceptibility tensor to $\chi_{ikk}^{(2)}$. This susceptibility can be further simplified if $\omega_0$ is far away from any resonance frequency of the material, and only OR effects are considered in $P_{NL}$, leaving the susceptibility $\chi^{OR}(\omega;\omega_0)$ as a function of only $\omega_0$ and the resultant THz polarisation frequency $\omega$.

Implementing these adjustments to the nonlinear wave equation and solving for the electric field gives:

\small
\begin{equation}\label{E}
    E(\omega,z)=\frac{\mu_0\chi^{OR}(\omega;\omega_0)\omega I_0(\omega)}{n(\omega_0)\{\frac{c}{\omega}[\frac{\alpha_T(\omega)}{2}+\alpha_0]+i[n(\omega)+n_g]\}}\frac{exp[-i\frac{\omega n(\omega)}{c}z]exp[-\frac{\alpha_T(\omega)}{2}z]-exp(-i\frac{\omega n_g}{c}z)exp(-\alpha_0z)}{\frac{\alpha_T(\omega)}{2}-\alpha_0+i\frac{\omega}{c}[n(\omega)-n_g]}
\end{equation}

\normalsize

Here, $I_0$ is the pump pulse spectrum that has transmitted through the incident interface of the crystal, $n(\omega)$ is the refractive index of the crystal at frequency $\omega$, $n_g$ is the group index of the pump pulse, $a_0$ amd $a_T$ are the absorption coefficient of the crystal at the pump pules and THz frequencies respectively, $c$ is the speed of light, and lastly $z$ is the propagation depth of the pump pulse.

The second fraction in Eq.\ref{E} is a complex entity with the unit of a length. The magnitude of this fraction is at its highest when the absorption coefficients $\alpha_T$ and $\alpha_0$ are low, and when the phase of the pump and the THz pulses match ($n(\omega)=n_g$). Therefore, the magnitude of this fraction can be defined as the effective generation length:

\begin{equation}
    L_{gen}(\omega,z)=\left(\frac{exp[-\alpha_T(\omega)z]+exp(-2\alpha_0z)-2exp\{-[\frac{\alpha_T(\omega)}{2}+\alpha_0]z\}cos\{\frac{\omega}{c}[n(\omega)-n_g]z\}}{[\frac{\alpha_T(\omega)}{2}-\alpha_0]^2+(\frac{\omega}{c})^2[n(\omega)-n_g]^2}\right)^{1/2}
\end{equation}

The various electromagnetic attributes of BNA, as was utilised in these calculations are as follows: refractive index in the THz range was obtained through Lorentz oscillator fitting of the experimental data, THz range absorption from the Lorentz oscillator parameters of Miyamoto Et Al. \cite{Miyamoto:09}, optical range refractive index from the Sellmeier equation parameters presented in \cite{Fujiwara_2007}, the optical range absorption was calculated from the power propagation model \cite{inf_model}. Lastly, the pump pulse was simulated as a Guassian distribution in frequency through matching the FWHM of the experimental pump pulse with the standard deviation of the distribution.

\section{Results}

\begin{figure}
    \centering
    \includegraphics[width=0.8\linewidth]{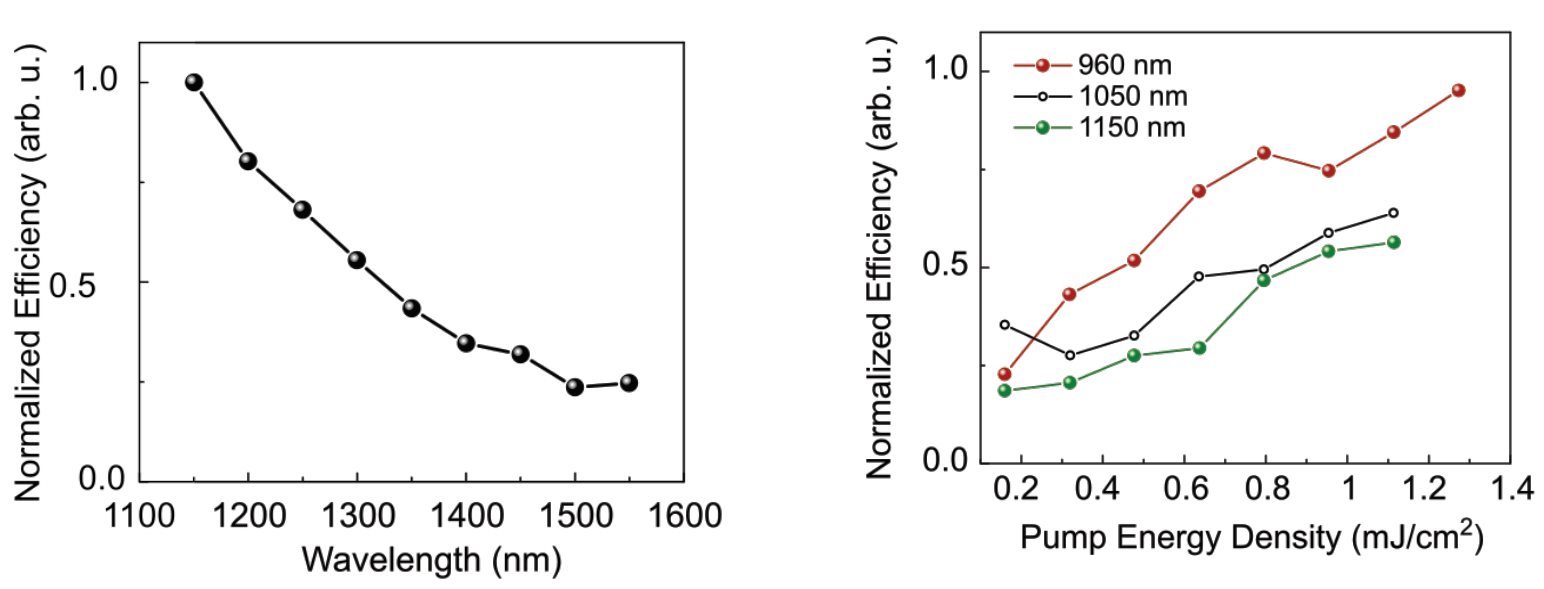}
    \caption{Normalised optical to THz conversion efficiency as a function of a) pump wavelength and b) fluence for a $550$ $ \mu m$ thick BNA crystal. As pump wavelength increases, a decrease in conversion efficiency is observed in a). In b), the highest efficiencies were achieved for the shortest wavelength at $960$ $nm$, reaching $0.6\%$ at $1.3$ $mJ/cm^2$. All three measurement sets suggest a linear regime; saturation of the conversion efficiency was not observed.}
    \label{efficiency}
\end{figure}
\begin{figure}
    \centering
    \includegraphics[width=0.8\linewidth]{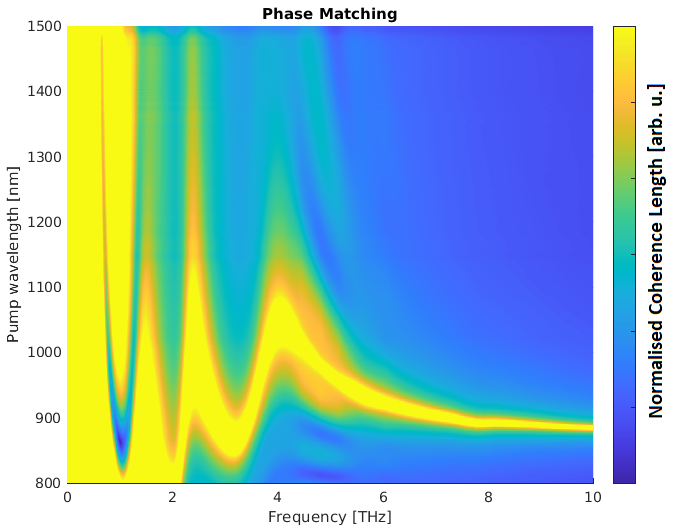}
    \caption{Calculated normalised coherence length with respect to the pump wavelength and generated THz frequency. We observe favourable phase matching between all pump wavelengths in the range of 0-1.2 THz, with an exception of a thin band around $1$ $THz$ between the pump wavelengths of 800-900 nm. The coherence length decreases at higher THz frequencies, leaving only a thin band near the pump wavelength of 900nm. The simulation confirms a broad spectrum and maximum coherence length for $\sim1$ $um$ driving pulses.}
    \label{phase_match}
\end{figure}


To demonstrate the optical to THz conversion capabilities of BNA, the THz frequency integrated amplitude of the generated spectra
are presented in Fig.\ref{efficiency}.a. For a crystal of thickness $550\ \mathrm{\mu m}$ under constant pump energy, a gradual decay of efficiency is observed as pump wavelength increases. 
Furthermore, Fig.\ref{efficiency}.b shows a linearly increasing trend in efficiency as pump energy density increases, with the pump wavelength tuned to $960$ $nm$, $1050$ $nm$, $1150$ $nm$. The highest efficiency was observed at $0.6 \%$ for the measurement of the shortest pump wavelength of $960$ $nm$ and the highest pump fluence of $1.3$ $mJ/cm^2$, without observable efficiency saturation.

Fig.\ref{phase_match} depicts the calculated normalised coherence length of BNA. For all pump wavelengths between $800-1500$ $nm$ a large coherence length was observed between $0-1.2$ $THz$, and decreases at higher $THz$ frequencies with a exception of a narrow band of large length $\sim 900$ $nm$. These results concur with those of Fig.\ref{efficiency}; THz generation efficiency is highest at $\sim 900$ $nm$, and decreases as pump wavelength increases.

\begin{figure}
    \centering
    \includegraphics[width=0.7\linewidth]{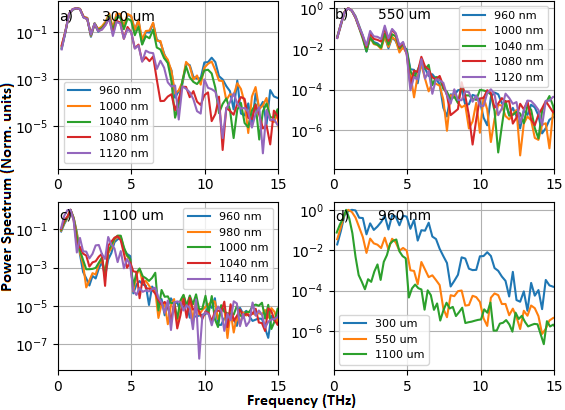}
    \caption{Logarithmic power spectra for crystals with different thicknesses pumped varying wavelengths. a) Crystal thickness of $300$ $\mu m$, b) $550$ $\mu m$ and c) $1100$ $ \mu m$. (d) Comparison of constant pump wavelength spectral contents for different crystal thicknesses.}
    \label{spec}
\end{figure}

The individual normalised THz spectra, under varying pump pulse wavelengths, are presented in Fig.\ref{spec} for crystal lengths of a) $300$ $\mu m$, b) $550$ $\mu m$ and c) $1100$ $\mu m$. A comparison between the spectra generated with different crystal thicknesses under the pump wavelength of $960$ $nm$ is shown in Fig.\ref{spec}.d. These results agree with Fig.\ref{phase_match}; high spectral density were observed between $1-2$ $THz$ for all pump wavelengths, with a decreasing trend of spectral density at higher $THz$ frequencies. 



\begin{figure}
    \centering
    \includegraphics[width=0.8\linewidth]{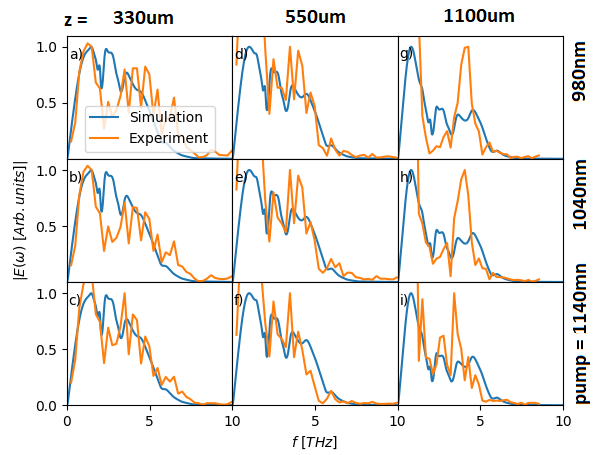}
    \caption{Comparison between experimental (arbitrary normalisation) and simulation results for BNA crystal thickness of $300$ $um$ (left column), $550$ $um$ (middle column) and $1100$ $um$ (right column) at pump wavelengths of $980$ $nm$ (top row), $1040$ $nm$ (middle row), $1140$ $nm$ (bottom row).}
    \label{comp}
\end{figure}

Lastly, a comparison between the experimental data and the calculated spectra at pump wavelengths of $980$, $1040$ and $1140$ $nm$ is shown in Fig.\ref{comp} for crystal lengths of a) $330$ $um$, b) $500$ $um$ and c) $1100$ $um$. All calculated spectra match the experimental data well at $1-2$ $THz$ frequencies, with an overestimate of a peak at $\sim 2.5$ $THz$ in the $300$ $um$ crystal. Moreover, a spectral peak was observed up to $4-5$ $THz$ in all data sets, this is in agreement with the large coherence length band at $4$ $THz$ spanning between $1000-1100$ $nm$ pump wavelengths. Finally, the broadest spectra (up towards $7$ $THz$) were observed with the thinnest crystal ($330$ $um$) at pump wavelengths closest to $900$ $nm$, coinciding with the calculated coherence length. 

\section{Conclusion}
In this work, we investigate THz generation of organic crystal BNA near ytterbium laser wavelength for different crystal thicknesses and pump wavelengths. We present the THz conversion efficiency, as well as the generated spectral content, to demonstrate a linear relationship between pump energy and efficiency. The highest optical to THz conversion efficiencies, along with a broad THz spectrum reaching up to 7 THz, can be found for driving pulses centred at around $1$ $\mu m$, as demonstrated both experimentally and theoretically. This paves the way for an additional application of the Ytterbium laser wavelength, and for BNA as a potential next generation THz source.

\section{Disclosures}
The authors declare no conflicts of interest.

\bibliography{sample}

\end{document}